\documentclass[11pt,b5paper]{article}
\usepackage{proc_dd}                             
\usepackage{graphicx}
\newcommand{\dnu}{\partial_n u}
\newcommand{\Do}{\Omega}
\newcommand{\Ga}{\Gamma}
\newcommand{\Neu}{\mathcal{N}}
\newcommand{\sym}{\sigma}
\newcommand{\imped}{\eta}
\newcommand{\Ss}{\mathbf{S}}
\newcommand{\Zz}{\mathbf{Z}}
\newcommand{\Rr}{\mathbf{R}}

\newcommand{\han}{H_{|n|}^{(1)}}
\newcommand{\dst}{\displaystyle}
\newcommand{\syml}{\sym_{\mathrm{lim}}}
\def\ve#1{\mathbf{#1}}
\begin{document}
\title{Symbol of the Dirichlet-to-Neumann operator in 2D diffraction
problems with large wavenumber}

\author{Margarita F.~Kondratieva, 
\textbf{Sergey Yu.~Sadov}}
{Department of Mathematics and Statistics, \\ 
Memorial University of  Newfoundland, Canada}%
{sergey@math.mun.ca}                              

\index{Kondratieva M.F.}                         
\index{Sadov S.Yu.}                              

\bigskip

\begin{abstract}
   Consider the Dirichlet-to-Neumann operator $\mathcal{N}$
   in the exterior problem for the 2D Helmholtz equation 
   outside a bounded domain with smooth boundary. 
   Using parametrization of the boundary by normalized arclength, 
   we treat $\mathcal{N}$ as a pseudodifferential operator on the unit 
   circle. We study its discrete symbol. 
   
   We put forward a conjecture on the universal behaviour,
   independent of shape and curvature of the boundary,
   of the symbol as the wavenumber $k\to\infty$. The conjecture is
   motivated by an explicit formula for circular boundary,
   and confirmed numerically for other shapes. It also agrees,
   on a physical level of rigor, with Kirchhoff's approximation.
    The conjecture, if true, opens new ways in numerical analysis
   of diffraction in the range of moderately high frequencies. 
     
\end{abstract}


\section{Introduction}
This work is a part of research aimed at an accurate and robust numerical
algorithm for diffraction problems in mid-high frequency range, where
the standard Boundary Integral Equation methods fail due to large matrix
size and, more importantly, to numerical contamination in quadratures.
A natural idea to use the knowledge of geometric phase and to separate
fast oscillations from slowly varying amplitudes 
has been converted to a practical method \cite{Ned1},\cite{Ned2}
with recent enhancements \cite{Darr}. A drawback of that approach
occurs in the presence of flattening boundary regions, where Kirchhoff's
amplitude becomes singular. From numerical analyst's point of view,
a method that has problem with {\it small} curvature is anti-intuitive.   

The point of our approach is to {\bf look for an object in theory whose
high-frequency asymptotics stands flattening well and isn't
sensitive to convexity assumptions}. We suggest that 
the symbol of Dirichlet-to-Neumann operator might be such an object.

We consider the 2D case and don't claim a ready-made extension
of our results in 3D. As a technical reason, we need a well defined 
{\it full} symbol of a pseudodifferential
operator on a compact manifold (the boundary). In the 2D case, 
the boundary is a closed curve, so a special version of 
the PDO theory with discrete frequency variable is applicable,
which attends to smooth kernels and doesn't require partitions of unity.    

\section{Dirichlet-to-Neumann operator}
Let $\Do$ be a bounded domain in $\Rr^2$ with smooth boundary $\Ga$.
The exterior Dirichlet problem for the Helmholtz equation
in polar coordinates $r$, $\phi\,$ reads \newpage
\begin{equation}
\label{prob}
 \begin{array}{l}
 \dst
  \Delta u\,\equiv\,
  \frac{\partial^2 u}{\partial r^2}\,+\,
  \frac{1}{r}\,\frac{\partial u}{\partial r}
  \,+\,\frac{1}{r^2}\,\frac{\partial^2 u}{\partial \phi^2}
  \,=\,-k^2 u 
  \quad\mbox{\rm in} \quad \Rr^2\setminus\bar{\Do},
  \\[2ex]
  \dst
  \frac{\partial u}{\partial r}\,-\,iku\,=\,o(r^{-1/2}),
  \quad r\to\infty,
  \\[2ex]
  u|_{\Ga}= f.    
 \end{array}
\end{equation}
For a function $\,f\in C^1(\Ga)$, the problem has unique solution $u$, and 
its normal derivative $\,g=\partial_n u|_{\Gamma}\,$ is a continuous 
function on $\Ga$. 
(For sharper conditions 
see e.g.\ \cite{Col1}, \cite{Ned3}.)
The map $\,\Neu:\, f\to g\,$ is called the {\it Dirichlet-to-Neumann operator} 
for Problem (\ref{prob}). 

\section{Operator $\Neu$ for the exterior of the unit disc}

Here the boundary is the unit circle $\Ss$ and it can be parametrized
by $\phi$.
Consider Fourier series of the $2\pi$-periodic functions 
$\,f(\phi)=u|_{\Ss}\,$
and $\,g(\phi)=\dnu|_{\Ss}\,$
$$
 f(\phi)=\sum_{n\in\Zz} \hat{f}(n) e^{in\phi},
 \qquad
 g(\phi)=\sum_{n\in\Zz} \hat{g}(n) e^{in\phi}.
$$
The Helmholtz equation has outgoing elementary solutions in the product form
$$
 e^{i n\phi} \,\han(kr),
$$
where $\han$ are Hankel functions \cite{Abr} (9.1.3).
The solution $u(r,\phi)$ can be represented as a linear combination
of the elementary solutions. Matching Fourier coefficients in
the boundary data, we find
$$
 \hat{g}(n) = \sym(n;\,k) \hat{f}(n),
$$  
where (cf. \cite{Abr} (9.1.27.4) ) 
\begin{equation}
\label{symud}
\sym(n;\,k) \,=\,k\;\frac{\partial_k\han(k)}{\han(k)} \,=\,
-k\,\frac{H_{|n|+1}^{(1)}(k)}{\han(k)} \,+\,|n|.
\end{equation}
The operator $\Neu$ can be written in a pseudodifferential 
fashion, the function $\sym$ being its {\it discrete symbol}
$\,$ (here dependence of $\sym$ on the wavenumber $k$ is 
irrelevant and is omitted)
\begin{equation}
\label{pdnud}
 \Neu f(\phi)=\sum_{n\in\Zz} \,\sym(n) \,\hat{f}(n)\, e^{in\phi}.
\end{equation}

\section{Asymptotics of the symbol}
For a fixed $k$ and $n\to\infty$, we derive from \cite{Abr} (9.3.1)
$$
 \frac{H_{n+1}^{(1)}(k)}{H_{n}^{(1)}(k)}\,\sim\,\frac{2n}{k},
$$
so (in full agreement with pseudodifferential calculus)
\begin{equation}
\label{asymnud}
 \sym(n;\,k)\sim -|n|, \qquad |n|\to\infty.
\end{equation}
\newpage\noindent
On the other hand, if $n$ is fixed, then by 
\cite{Abr} (9.2.3)
\begin{equation}
\label{asymkud}
 \sym(n;\,k)\sim ik, \qquad k\to\infty.
\end{equation}
The next result, in which the ratio
$t=n/k$ is fixed, interpolates between the above two special cases.
Since $n$ is an integer, the integral part function $[\cdot]$ is involved.  

\medskip\noindent
{\bf Lemma}. {\it For any fixed $\,t\geq 0\,$ and $\,n=n(k,t)=
[kt]$, 
}
\begin{equation}
\label{limsymud}
\lim\limits_{k\to\infty}
\frac{\sym(n;\,k)}{k}\;=\;\syml(t)\;\;
\lefteqn{\raisebox{1.5ex}{$\Delta$}}{=}\;\;
\left\{\;\begin{array}{lr}
i\sqrt{1-t^2}, & \quad{\rm if}\quad t\leq 1\\[1.3ex]
-\sqrt{t^2-1}, & \quad{\rm if}\quad t\geq 1.
\end{array}\right.
\end{equation}

\medskip\noindent
This fact can be derived laboriously using \cite{Abr} (9.3.37--46) 
and asymptotics for the Airy functions. Instead, we demonstrate a
simple argument, which quickly produces the formula in the case $t\neq 1$,
and can be converted to a formal proof.

Consider a recurrence for the ratios $\,\mu_{\nu}\,$ of Hankel functions
of orders ${\nu}+1$ and ${\nu}$ with fixed argument $k$. According
to \cite{Abr} (9.1.27.1), we have 
$$
  \mu_{\nu} + \mu_{{\nu}-1}^{-1}=2\, {\nu}/k,
$$
or in the explicit difference form,
\begin{equation}
\label{besdifeq}
  \mu_{\nu}-\mu_{{\nu}-1}\,=\,-\mu_{{\nu}-1}\,-\,\mu_{{\nu}-1}^{-1}\,+\,2t_{\nu},
  \qquad t_{\nu}={\nu}/k.
\end{equation}
The ratio $t_{\nu}$ varies slowly. Consider the difference equation
with ${\nu}\sim n$ and frozen $\,t_{\nu}=t_n=t$. 
It has two complex stationary solutions
\begin{equation}
\label{statsol}
 \mu^{\pm}=t\pm\sqrt{t^2-1}.
\end{equation}
The equation in variations for (\ref{besdifeq}) is
$$
\delta\mu_{\nu}-\delta\mu_{{\nu}-1}\,=\,(-1+\mu_{{\nu}-1}^{-2})\,\delta\mu_{{\nu}-1}.
$$
Therefore, for $0<t<1$ both solutions (\ref{statsol}) are asymptotically stable, while
for $t>1$ the solution $\mu^{+}$ is asymptotically stable, and $\mu^{-}$
unstable. A solution of the equation with frozen $t_{\nu}$ approaches its
limit exponentially fast, so the value of ${\nu}$ near $n$ 
doesn't change significantly
while the stabilization occurs. Since by (\ref{symud})
$$
 \sym(n;\,k)=k\,(-\mu_n +t),
$$
and the attractor $\mu^{+}$ is unique in the case $t>1$, we immediately  
obtain (\ref{limsymud}) in that case. In the case $\,0<t<1$,  
the solution $\mu_{\nu}$ approaches $\mu^{-}$ with negative
imaginary part, because  
the initial value $\,\mu_0\approx -i$, cf.\ (\ref{asymkud}).

\medskip\noindent{\bf Note}. 
The Lemma holds for $\,t=1\,$ due to the asymptotics derived from 
\cite{Abr} (9.3.31--34)
$$
 \frac{\partial_k H^{(1)}_k(k)}{H^{(1)}_k(k)}\,\sim\,6^{1/3}\;
 \frac{(1+i\sqrt{3})\,\Gamma(2/3)}{(1-i\sqrt{3})\,\Gamma(1/3)},
 \qquad k\to\infty.
$$
 
\section{Disc of arbitrary radius}
Let $\,u(r,\phi;\,R,k)\,$ be a solution of Problem (1) with 
wavenumber $k$ outside a circle of radius $R$. 
Then   
$\,u(r/R,\phi;\,1,kR)\,$ is a solution of Problem (1) with
wavenumber $kR$ outside the unit circle. The Dirichlet data for
the two functions (as functions of $\phi$) are identical,
$\,f_{R,k}(\phi)=f_{1,kR}(\phi)$. The Neumann data are related via
$$
  g_{R,k}(\phi) \,=\,\partial_r u(r,\phi;\,R,k)\left|_{r=R}
  \lefteqn{\phantom{0_0}}\right.
  \,=\, R^{-1}\, g_{1,kR}(\phi).
$$
Correspondingly, the symbol of the operator $\Neu$ for the disk of radius
$R$ is
\begin{equation}
\label{d2nR}
 \sym_{R}(n,\;k)\,=\,R^{-1}\,\sym_1(n,\;kR),
\end{equation}
so the limit formula (\ref{limsymud}) of Lemma holds 
with $\;n=n(k,t)=[kRt]$. Equivalently, we can write the argument of 
the limit function $\,\syml(t)\,$  as
\begin{equation}
\label{scalexi}
 t \,=\, \frac{n}{kR}\,=\, \frac{2\pi}{L}\,\frac{n}{k},
\end{equation}
where $L=2\pi R$ is the circumference of the boundary.
Notice that the factor $\;2\pi/L\,$ is the Jacobian 
$\,\partial\phi/\partial s$
of the boundary parameter change from the arclength $s$ to $\phi$.

\smallskip
In the limit $R\to\infty$ the disk becomes a half-plane and 
an analog of the asymptotic formula (\ref{limsymud})
is an {\it exact} formula (\ref{symnhp}) below. 
 
\section{Half-plane}
For the Helmholtz equation in the half-plane $\,(x\in\Rr,\; y>0)$, 
Sommerfeld's radiation condition is replaced by a condition that
explicitly specifies allowed harmonics in the decomposition of
any outgoing solution. Namely, 
two differently behaved families of elementary outgoing solutions 
are given by 
$$
w(x,y;\,\xi)\;=\;\left\{
\begin{array}{l}
\exp\{i x \xi\,+\,iy\sqrt{k^2-\xi^2}\},\qquad -1<\xi<1,
\\[2ex]
\exp (i x \xi) \,\exp(-y\sqrt{\xi^2-k^2}),\qquad |\xi|>1.
\end{array} 
\right.
$$
The general outgoing solution has the form
\begin{equation}
\label{plwaves}
 u(x,y)=\int_{-\infty}^{\infty} \hat{f}(\xi)\, w(x,y;\,\xi)\,d\xi,
\end{equation}
(we don't discuss possible classes to which the function
$\hat{f}(\xi)$ may belong).  
 
It is readily seen that $\hat{f}(\xi)$ is the Fourier
transform  of the Dirichlet boundary 
data $\,f(x)=u(x,0)$. 
Differentiating (\ref{plwaves}) with respect
to $y$, we obtain the Fourier representation for the Neumann data $g(x)$. 
The formula for the Dirichlet-to-Neumann operator, an analog of (\ref{pdnud}), reads 
$$
 \Neu f(x)\,=\,\frac{1}{2\pi}\,
 \int_{-\infty}^{\infty} \,\sym(\xi) \,\hat{f}(\xi) e^{ix\xi}\,d\xi,
$$
where the symbol $\,\sym(\xi)=\sym(\xi; k)\;$ is
$\;\partial_y w(x,y;\,\xi)/w(x,y;\,\xi)|_{y=0}$, i.e. 
\begin{equation}
\label{symnhp}
 \sym(\xi; k) \,=\, k\,\syml({\xi}/{k}).
\end{equation}

\section{Periodic pseudodifferential operators}

The reader can probably see what conclusion we are about to 
draw from the above examples. Let us complete
technical preparations, then formulate the main conjecture.   

\smallskip
Recall briefly and informally some basic notions regarding pseudodifferential
operators on the unit circle $\Ss$. See \cite{Agr}, \cite{Saranen}
for a full account of the topic.  

\smallskip
Let $a(\phi,n)$ be a function on $\Ss\times\Zz$, which
satisfies certain regularity conditions. The function $a(\phi,n)$
is the {\it discrete symbol} of the {\it periodic pseudodifferential
operator} (PPDO) $A\,$ defined by the formula
$$
 Af(\phi)\,=\,\sum_{n\in\Zz} a(\phi,n)\,\hat{f}(n) e^{in\phi}.
$$ 
Here $f(\phi)$ is a $2\pi$-periodic function and $\hat{f}(n)$ its
Fourier coefficients.

\smallskip
The symbol $\sym(n)$ introduced in (\ref{pdnud}) 
does not depend on $\phi$. Such symbols are called {\it constant symbols},
and corresponding operators are {\it shift invariant}. 

\smallskip 
The theory of PPDO applies not only to operators on the unit circle, but
to operators on any smooth closed curve, since functions on closed
curved can be identified with $2\pi$-periodic functions by reparametrization.

\smallskip
The symbol $a(\phi,n)$ 
of a PPDO $A$ typically has an asymptotic expansion in 
decreasing powers of $n$. The {\it principal symbol} 
is the leading term in the asymptotics
$$
 a(\phi,\pm|n|)\,=\, a_0^{\pm}(\phi)\, |n|^{\alpha}\,+\,o(|n|^{\alpha})
 ,\qquad |n|\to\infty,
$$
and $\,\alpha$ is the {\it order} of $A$. For example, for any domain
the operator $\Neu$ is a PPDO of order of 1, and for the unit disk its 
principal symbol is $-|n|$, cf.\ (\ref{asymnud}).

\smallskip
Theory of PPDO is somewhat simpler
than the general theory of pseudodifferential operators on compact manifolds
(see e.g. \cite{Taylor}). 
The definition of a general PDO uses partition of unity.
Only the principal symbol can be defined globally. 

The discrete symbol $\,a(\phi,n)\,$
of a {\it classical}\ PPDO agrees on $\Zz$ with a symbol $\tilde a(\phi,\xi)$ 
defined in the general theory, modulo a function with asymptotics 
$O(|n|^{-\infty})$. 

Reconstruction of an operator by its symbol
in the general theory assumes that operators with smooth kernels are
neglected. It isn't convenient when one studies double asymptotics 
(in $\xi$ and $k$), since the behaviour of the neglected part with respect 
to $k$ is not controlled. Correspondence between
operators and symbols in the theory of PPDO with discrete symbols
is strict and preserves full information in both directions.  
 
\section{Limit Shape Conjecture}

We return to Problem 1 with general boundary $\Gamma$. 
Denote the length of $\Ga$ by $L$. Let $s$ be the arclength parameter on $\Ga$ 
(with an arbitrarily chosen starting point), and set
$$
 \psi\,=\,s\,\frac{2\pi}{L}, \qquad 0\leq \psi<2\pi.
$$
Consider the operator $\Neu$ as a PPDO (with respect to the parametrization 
by $\psi$). 
Denote its symbol as $\,\sym_{\Ga}(\psi,n;\,k)$, emphasizing
dependence on the wavenumber $k$.

\newpage
\noindent
{\bf Conjecture}.
{\it For any fixed $\,t\in\Rr\,$ and $\,n=n(k,t)=
[(L/2\pi)kt]$, there exists
$$
\lim\limits_{k\to\infty}
\frac{\sym_{\Ga}(\psi,n;\,k)}{k}\;=\;\syml(t).
$$
uniformly w.r.t.\ $\psi$. The universal function $\,\syml(t)\,$ 
is defined in (\ref{limsymud}).
}

\medskip
\noindent
Let us say less formally:
$$
\sym_{\Ga}(\psi,n;\,k)\,\approx\, k\,\syml\left(\frac{2\pi}{L}\,\frac{n}{k}\right).
$$
We can make the conjecture even more readable at the expense of precise
mathematical meaning. Let us ignore problems associated with definition
of a global symbol of PDO in the standard theory, where the frequency
argument is continuous. Assume that $\;\sym_{\Ga}(s,\xi;\,k)\;$ is the
symbol of the operator $\Neu$ corresponding to the arclength parametrization
of the boundary. Then
\begin{equation}
\label{conj3}
\sym_{\Ga}(s,\xi;\,k)\;\approx\;\left\{\;\begin{array}{lr}
i\sqrt{k^2-\xi^2}, & \quad \xi<k\\[1.3ex]
-\sqrt{\xi^2-k^2}, & \quad \xi>k.
\end{array}\right.
\end{equation}
Thus the symbol for any boundary parametrized by the arclength is
asymptotically equal to the exact symbol for the half-plane. 
This conclusion is hardly surprising given that
at high frequencies the diffraction process is
well localized and (\ref{conj3}) takes place
for any disc --- see (\ref{d2nR}), (\ref{scalexi}) --- 
and doesn't refer to curvature. 

Our conjecture has no problems with tangent rays and shadow regions
since the formula doesn't depend on the boundary data. In particular
--- in the case of a plane incident wave --- the direction of
incidence has no effect on our claim.   
One can argue that the conjecture has no backing in the case of
non-convex scatterers. In that case it is supported by numerical results; 
see the last section of the paper.    

\section{Kirchhoff's approximation}
%

\noindent
\hbox to \textwidth{
\kern-4pt\parbox[b]{2.7in}
{
A relation between the boundary data $f$ and $g$ of an outgoing solution
can be described alternatively by the {\it impedance function}
$\;\imped=g/f$. It depends on the solution. 
However, according to Kirchhoff's approximation, at high frequencies
the impedance function approaches an universal function that depends
only on the boundary shape. Let us "derive" \linebreak
this approximation from
the Conjecture.  
}
\hfill
$\lefteqn{
{\mbox{
\includegraphics{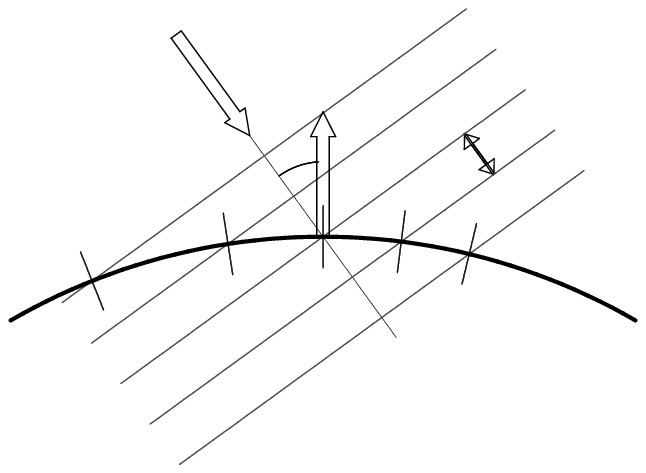}
}}}
\mbox{
\unitlength=1pt
\begin{picture}(180, 138)
\put(100,10){\bf Fig. 1}
\put(79,90){$\theta$}
\put(142,94){$\lambda$}
\put(41,110){$\ve{k_0}$}
\put(97,94){$\ve{n}$}
\put(80,51){$P$}
\put(175,51){$\Ga$}
\end{picture}
}$
}

\noindent
Consider an incident plane wave $u_{\rm inc}$ with the wave vector $k\ve{k_0}$,
$\,\|\ve{k_0}\|=1$.
Let $\ve{n}$ be the unit normal vector to the boundary
$\Ga$ at the given point $P\in\Ga$. Denote by $\theta$ the
angle between $\ve{k_0}$ and $\ve{n}\;$ (Fig.~1).
The incident wave length $\lambda=2\pi/k$ is the distance between
wave fronts with equal phases.  
The boundary value $\,u_{\rm inc}|_{\Ga}\,$ oscillates with
period $\Lambda=\lambda/\sin\theta$ near the point $P$.
We say that local frequency of  $\,u_{\rm inc}|_{\Ga}\,$
at $P$ is $\,\xi=2\pi/\Lambda=k\sin\theta$. 
Assuming Dirichlet's condition for the total field $\,u_{\rm inc}+u\,$
the boundary value $f=u_{\Ga}$ also oscillates with local frequency 
$\xi$ at $P$. 

From a physical point of view, 
the action of the operator $\Neu$ amounts to multiplication
of local Fourier harmonics by the values of the symbol $\sym_{\Ga}$
at corresponding space-frequency locations.
In the present case, where the harmonic with frequency $\xi$ dominates
at point $P$, formula (\ref{conj3}) implies
\begin{equation}
\label{kirh0}
\Neu u(P)\,\approx\, \sym_{\Ga}(P,\xi;\,k)\,u(P)\,\approx\,
i\sqrt{k^2-\xi^2}\,u(P)\,=\,ik\,\cos\theta\,u(P).
\end{equation}
Fig.~1 shows an illuminated region of the boundary, but the
argument holds for a shadow region as well. 
Formula (\ref{kirh0}) can be written in the form  
$$
 \imped(P)\approx ik\,\left|\langle\ve{k_0},\ve{n}(P)\rangle\right|,
$$
which is the classical Kirchhoff approximation formula \cite{Kirh}.
A rigorous mathematical treatment of Kirchhoff's approximation
(for convex domains) is given in \cite[Ch.~X]{Taylor}.

\section{Insufficiency of the naive local frequency analysis}

\medskip
\noindent
\hbox to \textwidth{
\kern-4pt\parbox[b]{2.7in}
{
The simplistic understanding of the symbol via local frequencies
fails in the following example.
Consider the horseshoe domain $\Do$ as shown on Fig.~2.
Let two solutions $u^{(1)}$ and $u^{(2)}$ of Problem (\ref{prob})
be defined outside $\Do$ as cylindrical waves
generated by the fictitious sources at the points $S_j$, $j=1,2$, inside $\Do$.
From asymptotics of Hankel's function
$\,H^{(1)}_0(kr)\,$ we see that 
if $k|S_j P|\gg 1$, then the 
two solutions yield opposite impedances $\,\eta_1(P)\approx -\eta_2(P)
\approx -ik$.
The local tangential frequency at $P$ 
is close to $0$ for both
solutions. Thus it is 
}
\hfill
\raisebox{30pt}{$\lefteqn{
{\mbox{
\includegraphics{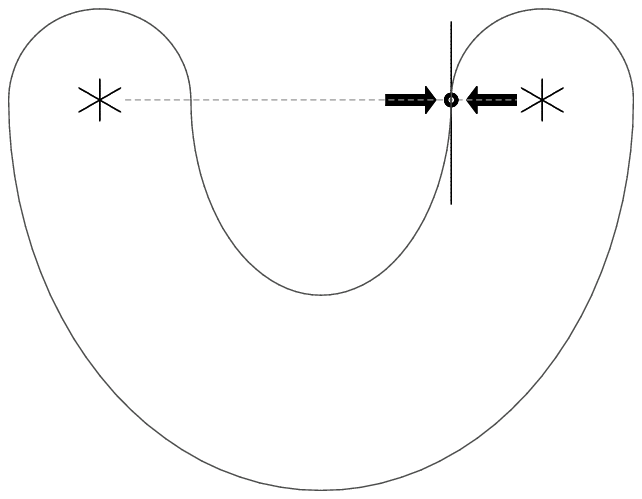}
}}}
\mbox{
\unitlength=1pt
\begin{picture}(181, 139)
\put(80,-15){\bf Fig. 2}
\put(31,94){$S_1$}
\put(142,97){$S_2$}
\put(117,127){$P$}
\put(75,40){$\Do$}
\end{picture}
}$}
}
%
impossible to determine the value 
$\,\sym_{\Ga}(P,0;\,k)\,$ consistently by this approach.

\section{Numerical verification of the Conjecture}
The following algorithm has been used to retrieve the symbol
of the operator $\,\Neu$. \linebreak
Assume $k$ is given. The algorithm
has three free parameters: number of nodes $N$ (taken in
the form $N=2^m$ for convenience), and coordinates $(x_S,\,y_S)$
of a fictitious source {\it inside} the domain $\Do$.

\newpage\noindent
{\bf Algorithm.}

\medskip\noindent
1. Find an equidistant partition of $\Ga$ by $N$
nodes $P_i$.

\medskip\noindent
2. Boundary data will be taken from the sample outgoing solution 
$$
 u(P)=H_0^{(1)}(k|PS|), \qquad P\notin\Do,
$$ 
where $S=(x_S,\,y_S)$ is the "source", and $P$ is an observation point. 
Compute the boundary data $\;f_i=u(P_i)$, $\;g_i=\dnu(P_i)$,
$\;i=1,\dots,N$.

\medskip\noindent
3. Compute discrete Fourier transforms $\hat{f}(n)$, 
$\hat{g}(n)$, $\,n=0,\dots,N-1$, of the arrays
$\{f_i\}$, $\{g_i\}$ using FFT algorithm. 
Only the first $n_{\rm max}$ Fourier coefficients
are considered reliable and are used in the sequel.

\medskip\noindent
4. Find the truncated symbol of a shift-invariant operator 
that takes $f$ to $g$:
$$
 \tilde\sym(n)\;=\;{\hat{g}(n)}/{\hat{f}(n)},
 \qquad n=0,\dots,n_{\rm max}-1.
$$     

\noindent
5. To verify the Conjecture, compare the values $\,k^{-1}\,\tilde\sym(n)\;$ 
to $\;\syml(2\pi n/kL)$, where $L$ is the length of $\Ga$.

\bigskip\noindent
We present results obtained for the {\it kite} domain 
\cite[p.~70]{Col2} shown on Fig.~3 and defined by the parametric equations
$$
 x(t)\,=\,\cos t\,+\,0.65 \cos 2t\,-\,0.65,
 \qquad
 y(t)\,=\,1.5\,\sin t,
 \qquad t=0\dots 2\pi.
$$  

$\qquad\lefteqn{
{\mbox{
\includegraphics{kite01.eps}
}}}
\mbox{
\unitlength=1pt
\begin{picture}(287, 216)
\put(90,0){{\bf Fig.~3}: Test domain ("kite")}
\put(110,144){$S$}
\end{picture}
}$

\medskip
\bigskip
\noindent
The parameters are: $\;k=200$, $\;N=2^{20}$, $\;S(-.7,\, .5)$. 
The width of the triangle on Fig.~3 is equal to 10 wavelengths.
In this example, length $L=9.32402$ and $\,kL/2\pi \approx297$.

On Fig.~4, the horizontal coordinate is  $t=2\pi n/kL$. Thick
lines show the normalized real (a), with negative sign,
and imaginary (b) parts
of the computed approximate symbol, $k^{-1}\tilde\sym(n)$.
Thin lines are the conjectured limit shapes.
The true symbol $\sym_{\Ga}$ in this case is non-constant, so the approximation
by a shift-invariant symbol depends on the chosen position of the source.
For a source closer to the center of the kite, 
oscillations near $t=1$ become smaller. However, in that case the 
computed values near $t=2$ oscillate wildly, because corresponding Fourier 
coefficients $\hat{f}(n)$ become evanescent.   

\bigskip
\noindent
$\lefteqn{
{\mbox{
\includegraphics{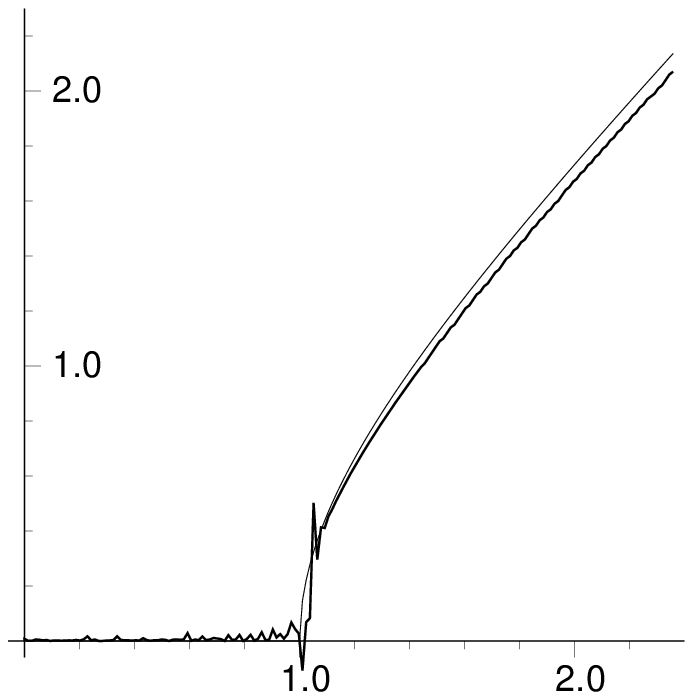}
}}}
\mbox{
\unitlength=1pt
\begin{picture}(198, 198)
\put(90,173){{\bf (a)}: $-$Re}
\put(184,21){$t$}
\put(70,-20){{\bf Fig.~4}: Computed symbol 
$\dst\;k^{-1}\tilde\sym(n)\;$ vs $\;\syml(t)$, $\;\;t=\frac{2\pi}{kL}n$}
\end{picture}
}$
$\lefteqn{
{\mbox{
\includegraphics{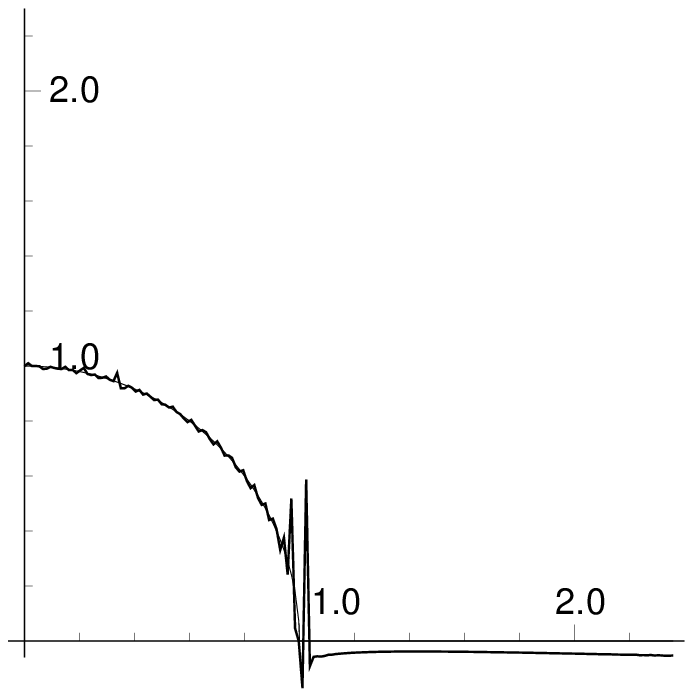}
}}}
\mbox{
\unitlength=1pt
\begin{picture}(197.5, 198)
\put(90,173){{\bf (b)}: Im}
\put(188,21){$t$}
\end{picture}
}$

\vspace{3\bigskipamount}
\noindent
The upper bound $t_{\rm max}\approx 2.3$ on the graphs corresponds
to $n_{\rm max}=700$ set in the computer program. Stabilization
of the Fourier coefficients at the upper end of this range occurs 
for the order of discretization $N\geq 2^{18}$. Obtaining stable 
values of the approximate symbol at larger values of $t$ requires use of 
larger values of $N$ that grow, roughly, exponentially with $t$.  
 
A program used for these calculations had a 12 byte long type 
for floating point operations ({\tt long double} in C). 
The results obtained with a 8 byte long arithmetics (C's type {\tt double})
were nearly identical. So in the considered example numerical errors
due to a limited precision are not an issue.
    
\section{Conclusion}
The main result is the proposed Limit Shape Formula (\ref{conj3})
for the symbol of the Dirichlet-to-Neumann operator for 
the standard 2D diffraction problem (\ref{prob}) with smooth
boundary. This asymptotics is independent
of the 
boundary data, of the boundary curvature, and of convexity 
assumptions. The limit function $\syml(t)$ defined in (\ref{limsymud})
varies slowly in its argument $t\sim {\rm const}\,n/k$, except near $t=1$. 
These features make the approximation  (\ref{conj3}) useful
for numerical completion of the boundary data set $(u|_{\Ga},\,\dnu|_{\Ga})$,
which yields the solution $u$ and the radiation pattern by Green's formula.
This approach includes and supersedes the classical Kirchhoff approximation.
We believe that the asymptotics can be enhanced and next, 
curvature-dependent, term(s) can be found from the theory
of pseudodifferential operators. In the especially important region, a 
narrow neighbourhood of $t=1$, methods for a field near a caustic
\cite{Babich} can be used.

\section{Acknowledgements}
We appreciate a friendly and fruitful atmosphere of the ~DD'03 ~Conference.
We \linebreak 
especially thank Prof.\ V.M.\ Babich for stimulating conversations.

This research was supported in part by a grant to M.K.\ from the
Natural Sciences and Engineering Research Council of Canada. 


\end{document}